# Absorption and Emission Dependences on Defect in CLC

A.H. Gevorgyan[1], K. B. Oganesyan[2]
Yuri Rostovtsev[3] , G. Kurizki[4] , Marlan Scully[5]

[1]Far Eastern Federal University, 10 Ajax Bay, Russky Island, Vladivostok 690922, Russia
[2]IAlikhanyan National Science Lab, Yerevan Physics Institute, Alikhanyan Br.2, 036, Yerevan, Armenia, bsk@.yerphi.am
[3]University of North Texas, Denton, TX, USA
[4]Weizmann Institute of Science, Rehotot, Israel
[5]Texas A&M University, College Station, Texas, USA

The influence of the defect position on absorption and emission in the cholesteric liquid crystal with an isotropic defect inside is studied. It is shown that for non-diffracting circularly polarized incident light absorption/emission is maximum if the defect is in the centre of the system; and for diffracting circularly polarized incident light absorption/emission is maximum if the defect is shifted from the centre of the system to its left border from where light is incident. The influence of anisotropic absorption in the cholesteric liquid crystal layer on photonic states of density was investigated, too.

## 1. Introduction.

Absorption and emission are not inherent properties of a material; rather, they arise due to interaction between the material and its local electromagnetic environment. Changing the environment can alter the absorption/emission rate. And we can change the electromagnetic environment by changing the material structure, as well.

Tailoring light absorption and emission with use of engineered nanophotonic structures (photonic crystals, metamaterials and etc) is an active area of research due to its potential applications, as in: miniature lasers and light emitting diodes [1-3]; single-photon generation; quantum information [4-7]; accumulation of solar energy [8-11]; sensing; integrated photonics, etc.

Absorption and emission are the very reliable methods of tuning of electromagnetic waves in a medium. Photonic crystals (PCs) and metamaterials possess interesting properties of absorption and emission. Particularly, here the suppression of absorption/emission effects takes place in the photonic band gap (PBG).

Suppression of absorption/emission outside the PBG occurs also for anisotropic absorption. An anomalously strong/weak absorption/emission outside the PBG nearby its borders takes place here, too.

Also, metamaterials with a few periods and large effect of anisotropy provide as much absorption/emission as usual PC layers with much more number of periods.

Absorption/emission of chiral PCs also has polarization peculiarities.

To increase absorption, it is necessary to obtain large light localization (accumulation) in as small a volume as possible.

One mechanism of large light accumulation is through multiple reflections – including diffraction reflections – from the system borders or from the inhomogeneities of the medium. And the presence of a defect in the PC structure provides large light accumulation at the defect mode [12].

In works [12-14] a new mechanism of light accumulation in the system, to wit, the diode mechanism of accumulation is discovered: if the directions of greater of two optical diodes are towards to each other (→ ←) the intensity $I$ on the surface of joining of these layers is greater than in the case when they directed from each other (← →). This, in its turn, implies the existence of the new mechanism of light accumulation specified by the nonreciprocal properties of system elements. In these works, the use of accumulated light energy, in particular, by way of isotropic absorption layer placed between two optical diodes or between the diode and the mirror is suggested.

Another system where large light accumulation takes place is plasmonic metal nanoparticles.

Cholesteric liquid crystals (CLCs) are the most widespread representatives of 1D chiral photonic crystals (CPCs) due to the possibility of spontaneous self-organization of their periodic structure and controllability of their PBG in a wide frequency range. The CLC parameters can be varied by an external electric/magnetic field, temperature gradient, UV radiation, etc.

CLCs also have some other surprising optical properties. CLCs with defects of various types in their structure have been considered for a few decades from the viewpoint of generating additional resonant modes in CLCs being investigated their possibilities of low-threshold lasing on these modes, their high rate absorption/emission possibilities, etc ([10-22] and [33-44]; see also references cited therein).

In this paper we investigated some new peculiarities of absorption/emission of a CLC layer with isotopic defect layer inside (Fig. 1).

## 2. Method of Analysis

The problem is solved by Ambartsumian's modified layer addition method adjusted to the solution of such problems (see [10-12]). A CLC layer with a defect inside (with a DL) can be treated as a multilayer system: *CLC(1)–[DL]–CLC(2)* (Fig. 1).

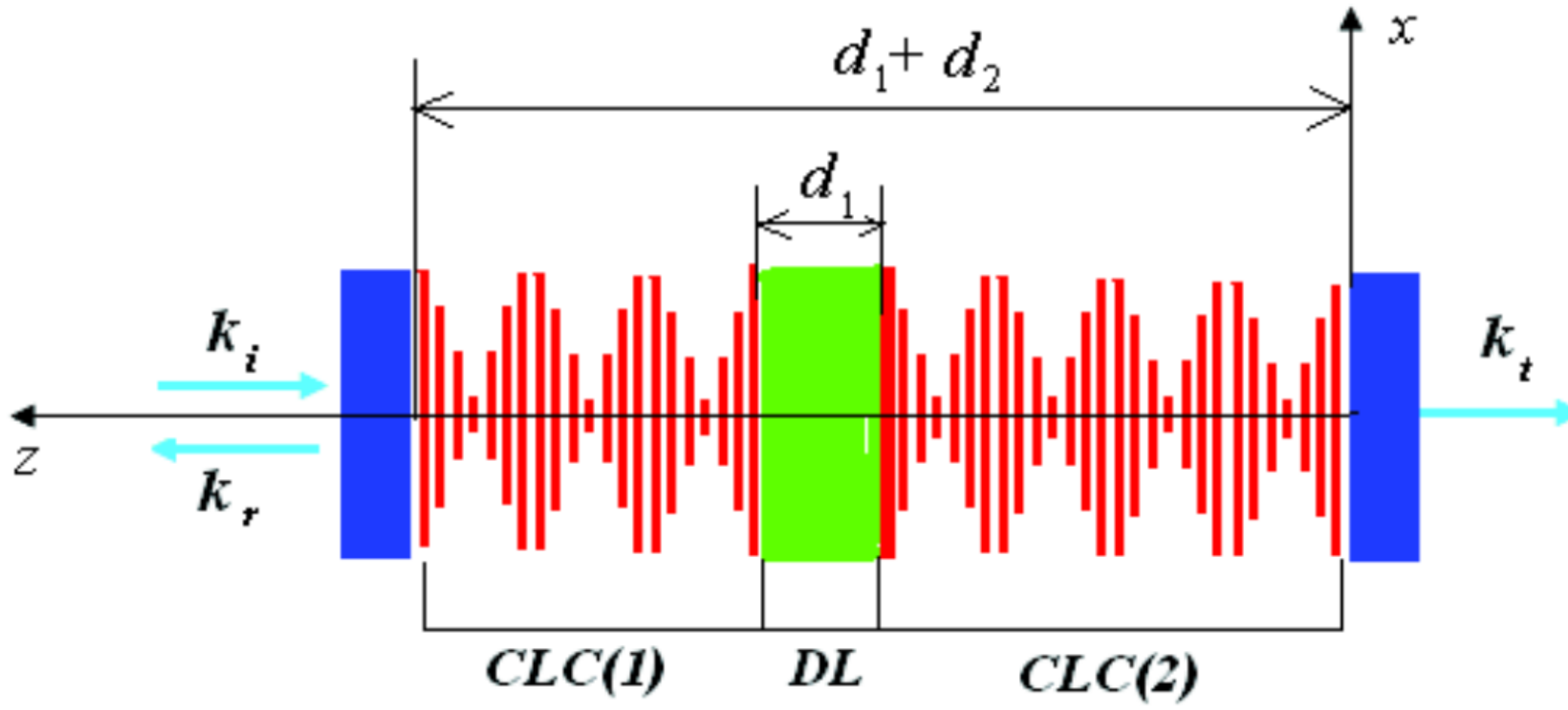

Fig. 1. (Color online) A sketch diagram of a modeled CLC cell with a defect layer inside.

According to Ambartsumian's modified layer addition method, if there is a system consisting of two adjacent (from left to right) layers, $A$ and $B$, then the reflection/transmission matrices of the system $A$+$B$, viz. $\hat{R}_{A+B}$ and $\hat{T}_{A+B}$, are determined in terms of similar matrices of its component layers by the matrix equations:

$$\hat{R}_{A+B} = \hat{R}_A + \tilde{\hat{T}}_A \hat{R}_B \left[ \hat{I} - \tilde{\hat{R}}_A \hat{R}_B \right]^{-1} \hat{T}_A,$$
$$\hat{T}_{A+B} = \hat{T}_B \left[ \hat{I} - \tilde{\hat{R}}_A \hat{R}_B \right]^{-1} \hat{T}_A, \qquad (1)$$

where the tilde denotes the corresponding reflection and transmission matrices for the reverse direction of light propagation, and $\hat{I}$ is the unit matrix. The exact reflection and transmission matrices for a finite CLC layer (at normal light incidence) and for a defect (isotropic) layer are well known [23, 24]. First, we attach the DL to the CLC layer (2) on the left side, using matrix equations (1). In the second stage, we attach the CLC layer (1) to the obtained DL–CLC layer (2) system, again on the left side.

The ordinary and extraordinary refractive indices of the CLC layers are taken to be $n_o = \sqrt{\varepsilon_2} = 1.4639$ and $n_e = \sqrt{\varepsilon_1} = 1.5133$; $\varepsilon_1$, $\varepsilon_2$ are the principal values of the CLC local dielectric tensor. The CLC layer helix is right handed and its pitch is: $p$ = 420 nm. These are the parameters of the CLC cholesteryl-nonanoate–cholesteryl-chloride–cholesteryl-acetate (20:15:6) composition at the temperature $t$ = 25°C. Hence, the light normally incident onto a single CLC layer – with right circular polarization (RCP) – has a PBG (which is in the range of λ ~ [614.8 – 635.6] nm), and the light with left circular polarization (LCP) does not have any. The isotropic defect layer refractive index was taken to be $n$ =1.8. We consider the case when our system is surrounded by the media with the reflection index $n_s$ on its both sides, which is equal to the CLC average reflection index $n_m = \sqrt{(\varepsilon_1 + \varepsilon_2)/2}$.

## 3. Results and Discussion

The presence of a thin defect in the CLC structure is known to initiate a defect mode in the PBG. This mode manifests itself as a dip in the reflection spectrum for right-handed circularly polarized light (the diffracting circular polarization) and as a peak in the reflection spectrum for left-handed circularly polarized light (the non-diffracting circular polarization). The defect mode is of either donor or acceptor type, depending on the optical thickness of the defect layer: the defect-mode wavelength increases from the minimum to the maximum of the band gap with an increase in the optical thickness of the defect layer; in this case, two defect modes arise near the both gap edges. With a further increase in the optical thickness of the defect layer, the long-wavelength mode vanishes, and the short-wavelength mode becomes red-shifted [10, 15]. In the case of a thick defect layer the number of the defect modes increases (see [19]). The frequency position of these modes and the number of excited modes can be varied by changing the thickness of the defect layer.

First we investigated the influence of the defect layer position on the absorption and emission peculiarities. Let the CLC and the isotropic defect layer be doped with dye molecules. Then this system is an amplifier in the presence of a pumping wave, i.e. we are discussing a planar resonator with active elements. The presence of the dye leads to the change of local refraction indices of the system sublayers. In this case, in the presence of the pumping wave, the effective imaginary parts of the local refraction indices in the CLC ($n''_{o,e}$) and that of the isotropic layers ($n''$) are negative – $n_{o,e} = \sqrt{\varepsilon_{1,2}} = n'_{o,e} + jn''_{o,e}$, $n = \sqrt{\varepsilon} = n' + jn''$ – and they all characterize the gain. In the case of the absence of the pumping wave (when the imaginary parts of the local CLC indices $n''_{o,e}$ and that of the isotropic layers $n''$ are positive) the quantity $A$=1–($R$+$T$), as mentioned above, characterizes the light energy absorbed by the system, while, in the case of gain, the value $|A|$ characterizes the radiation emitted by the system (we assume that the incident light intensity is unit, $I_0$ = 1).

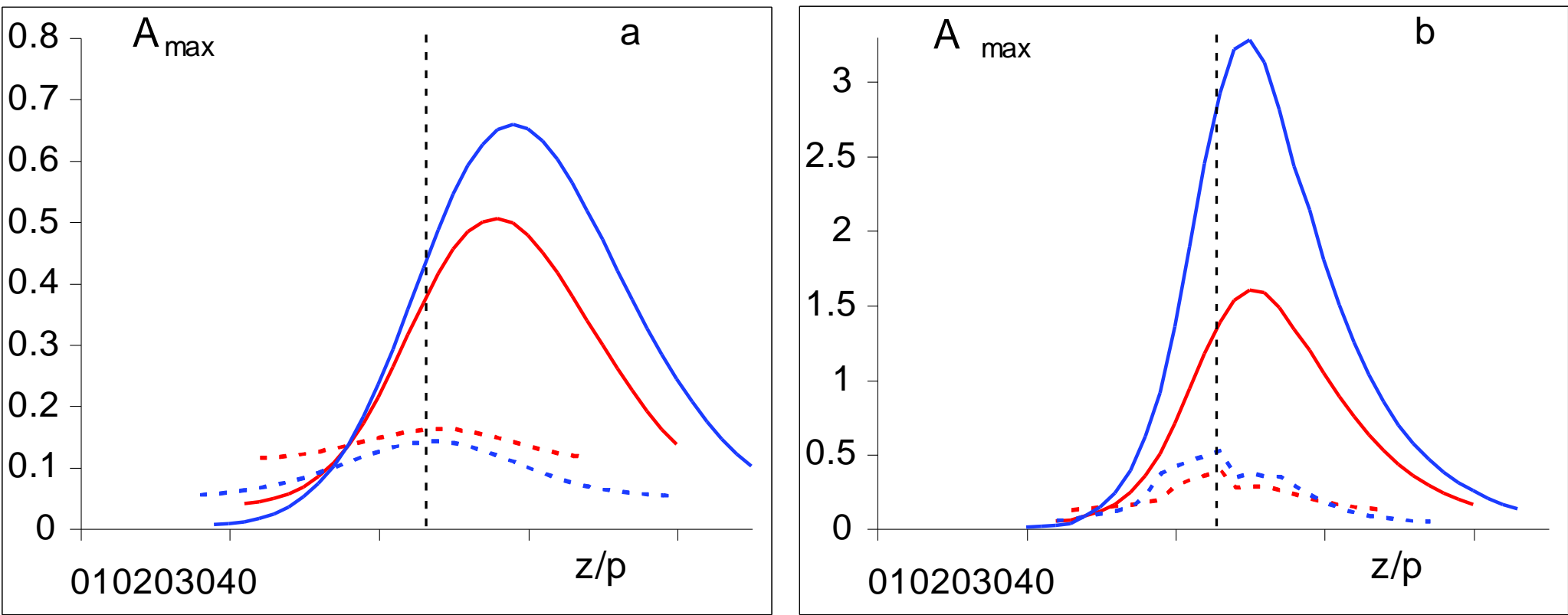

Fig. 2. (Color online) The dependences of (*a*) the maximum absorption $A_{\max}$ and (*b*) the maximum emission $|A_{\max}|$ at defect mode versus the defect layer position (*z/p*). The blue or dark gray lines correspond to $n''_o = n''_e = 0$ and $n'' \neq 0$. The red or light gray lines correspond to $n''_o = n''_e \neq 0$ and $n'' = 0$. The solid lines correspond to diffracting eigen polarizations and the dashed lines to non-diffracting ones.

In Fig. 2 the dependences of (*a*) the maximum absorption $A_{\max}$ and (*b*) the maximum emission $|A_{\max}|$ at the defect mode versus the defect layer position (*z/p*) are presented. The blue or dark gray line corresponds to $n''_o = n''_e = 0$ and $n'' \neq 0$, that is, for the case if absorption is absent in the CLC sublayers and it exists in the defect layer. The red or light gray lines correspond to $n''_o = n''_e \neq 0$ and $n'' = 0$, that is, for the case if absorption exists in the CLC sublayers and it is absent in the defect layer. The solid lines correspond to diffracting eigen polarizations and the dashed lines to non diffracting ones. (Eigen polarizations are the two polarizations of the incident wave that do not change when passing through the system.) For the single CLC layer the two eigen polarizations practically coincide with the two circular (right and left) polarizations.

The presence of the defect leads to changes of the eigen polarizations and these changes are significant at the defect modes. Due to these changes, polarization conversion at the defect layer borders takes place, which is conditioned by the dependence of reflection on polarization. This leads to an increase of reflection at the defect mode of the light with non-diffracting circular polarization.

One extreme case is important, namely, the case for which the defect layer refraction index and that of systems surrounding medium (on its both sides) coincide with the average refractive index of the CLC layer. In this case the eigen polarizations are again converted into two circular polarizations, and reflection of the non-diffracting circular polarization at the defect mode is absent (see [19] for more details).

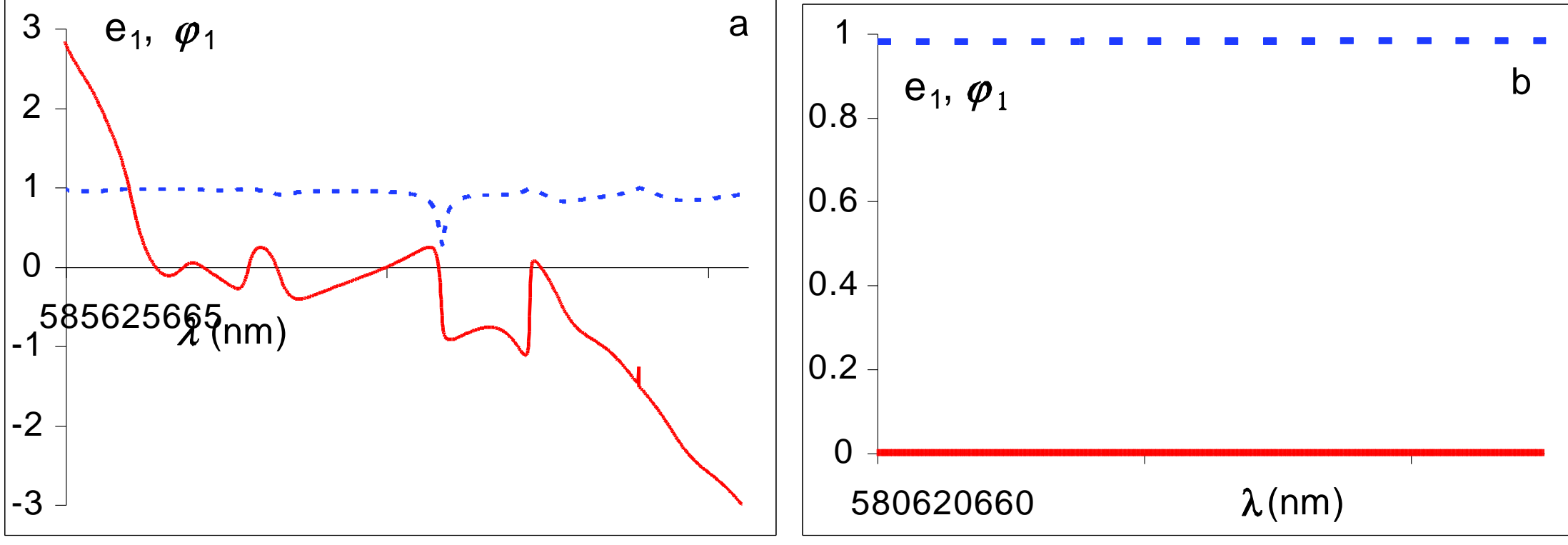

Fig. 3. (Color online) Spectra of ellipticity $e_1$ (dashed lines) and azimuth $\varphi_1$ (solid lines). (*a*) *n*=1.8, $n_s = n_m$; (b) $n = n_s = n_m$.

In Fig. 3*a,b* the ellipticity and azimuth spectra for the first eigen polarization are presented, for the two above-said cases. And we have $e_2 = -e_1$, $\varphi_2 = -\varphi_1$ for the second eigen polarization.

Now let us come back to Fig. 2. As it is seen in it, absorption and emission are minimal at the defect mode if the defect is shifted to the CLC layer borders. Absorption and emission are maximum for the incident light with non-diffracting circular polarization if the defect is in the centre of the system (the vertical dashed line indicates the centre of the system). For the incident light with diffracting circular polarization absorption and emission are maximal if the defect is not in the centre of the system, but is a little displaced from it to the left border of the system (from where light is falling). These curves also possess certain asymmetry with respect to their maximum values.

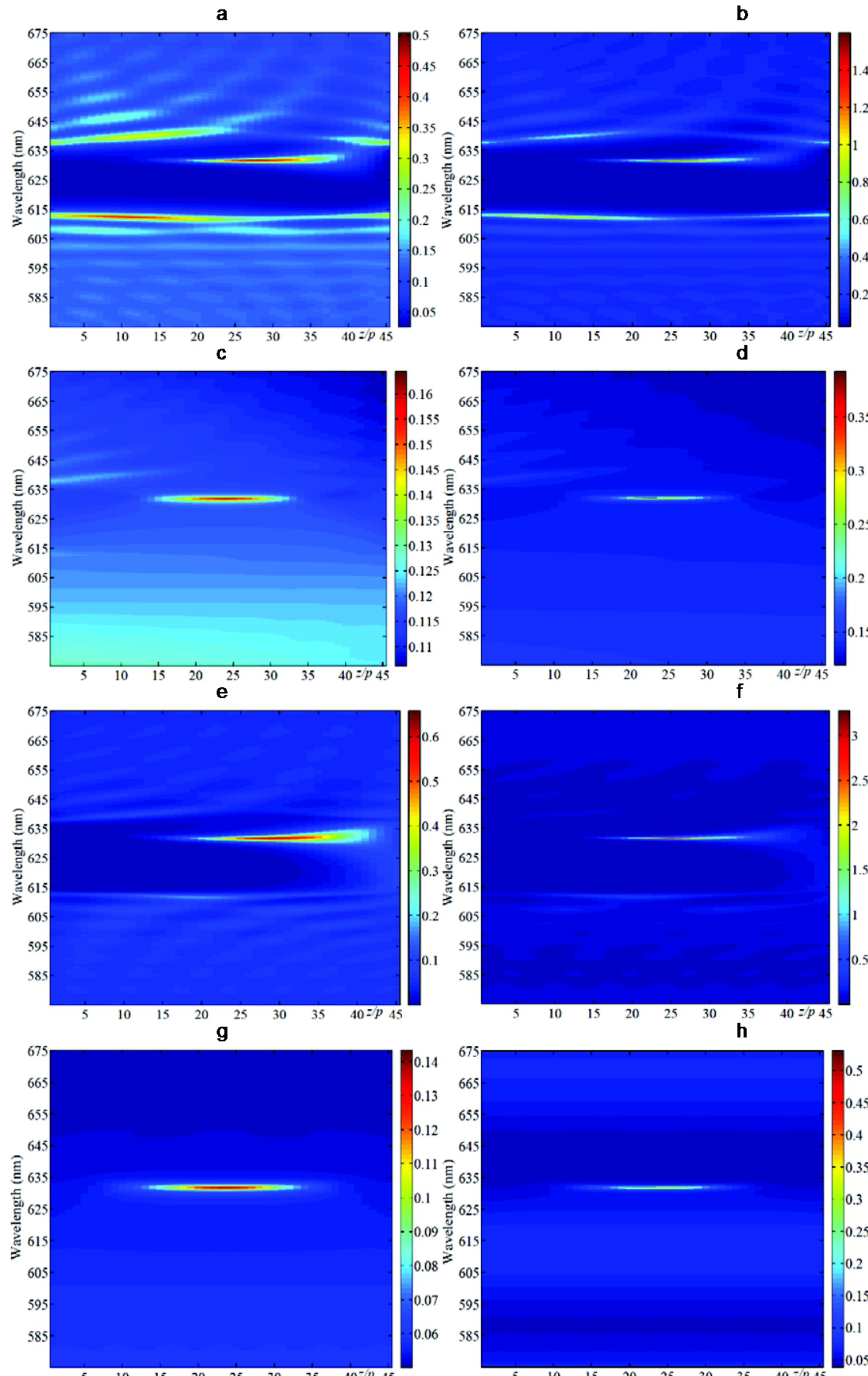

Fig. 4. (Color online) The evolution (*a,c,e,g*) of the absorption spectra and (*b,d,f,h*) of the emission spectra when the defect layer position changes. (*a,b,e,f*) are for the diffracting and (*c,d,g,h*) for non-diffracting circular polarizations. (*a,b,c,d*) correspond to case if $n_o'' = n_e'' \neq 0$ and $n'' = 0$, and (*e,f,g,h*) correspond to the case if $n_o'' = n_e'' = 0$ and $n'' \neq 0$.

In Fig. 4 the evolution (*a,c,e,g*) of the absorption spectra and (*b,d,f,h*) of the emission spectra when the defect layer position changes are presented for (*a,b,e,f*) the diffracting and (*c,d,g,h*) non-diffracting circular polarizations. (*a,b,c,d*) correspond to case if $n_o^{''} = n_e^{''} \neq 0$ and $n^{''} = 0$, and (*e,f,g,h*) correspond to the case if $n_o^{''} = n_e^{''} = 0$ and $n^{''} \neq 0$. Comparison of the figures shows that for the case of absorption the spectral line width for absorption is more than that for the gain spectral line width of emission.

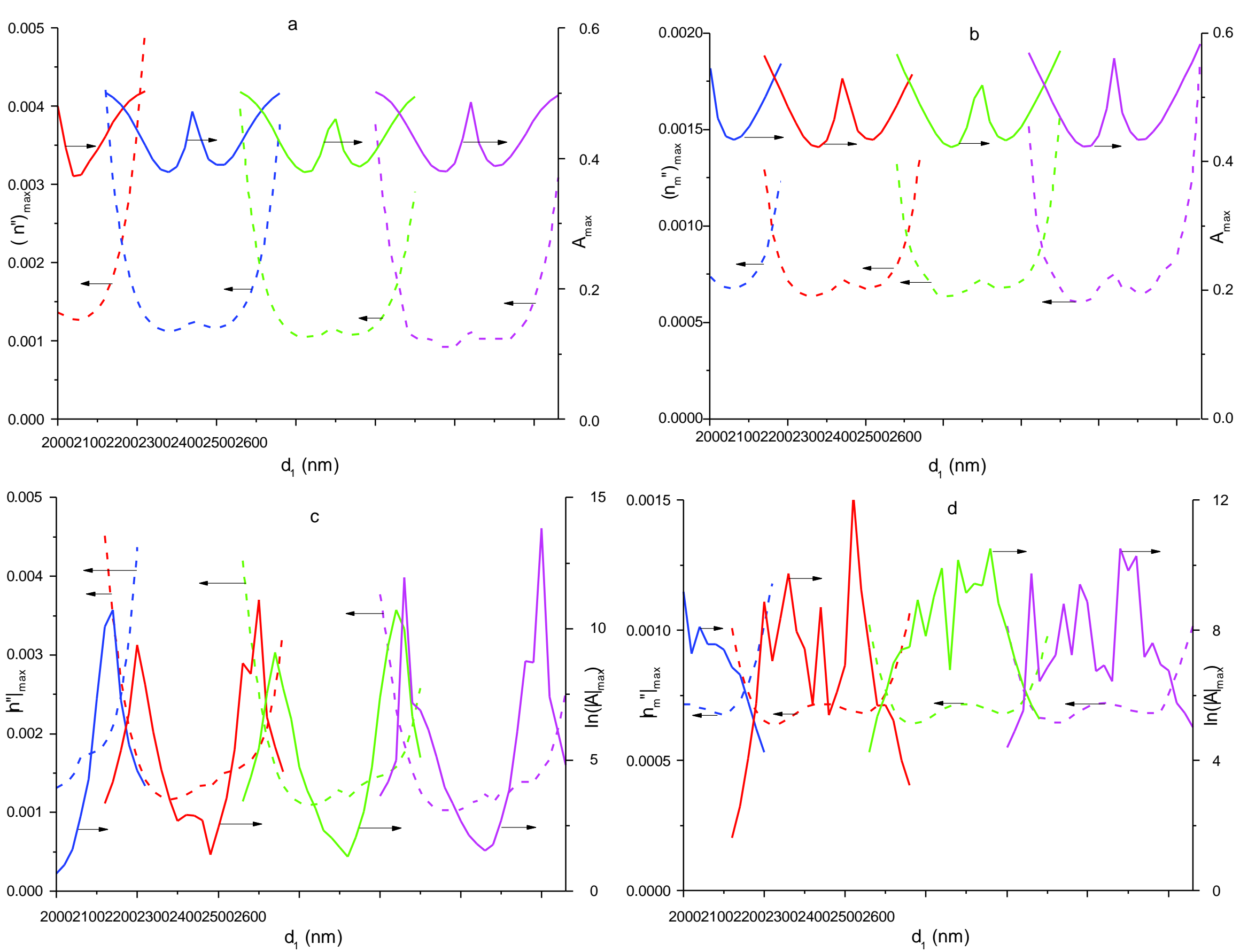

Fig. 5. (Color online) The dependences of (*a*) $(n^{''})_{\max}$ and $A_{\max}$, (*b*) $(n_m^{''})_{\max}$ and $A_{\max}$, (*c*) $\left|n^{''}\right|_{\max}$ and $\ln\left|A\right|_{\max}$ and (*d*) $\left|n_m^{''}\right|_{\max}$ and $\ln\left|A\right|_{\max}$ versus the defect layer thickness $d_1$.

Furthermore, we investigated the absorption and emission dependences on the parameters characterizing the absorption and amplification for various thicknesses of the defect layer. The dependences of absorption and emission versus the above said parameters have maxima. In Fig. 5 the dependences of $(n^{''})_{\max}$, $(n_m^{''})_{\max}$, $A_{\max}$, $\left|n^{''}\right|_{\max}$, $\left|n_m^{''}\right|_{\max}$ and $\ln\left|A\right|_{\max}$ on the defect layer thickness are presented. $A_{\max}$ is the maximum value of the absorption $A$ if the parameter characterizing absorption is increasing (i.e. versus $n^{''}$, if the defect layer is the absorbent and $n_m^{''} = 0$ or $n_m^{''}$, if the CLC layer is the absorbent and $n^{''} = 0$). Analogically, $\ln\left|A\right|_{\max}$ is the maximum emission logarithm versus the parameter characterizing the gain of the medium (i.e. either $\left|n^{''}\right|$, if the defect layer is the amplifier or $\left|n_m^{''}\right|$, if the CLC layer is the amplifier). $(n^{''})_{\max}$ is the value of $n''$ for which $A = A_{\max}$ if $n''$ increases. The following quantities, $(n''_m)_{\max}$, $|n''|_{\max}$ and $|n''_m|_{\max}$ are defined in the same way. The curves of different colors correspond to different defect modes (if the thickness of the defect layer is increased new defect modes appear, as was said above). As is seen in Fig. 5*a,* $(n'')_{\max}$ is sharply decreased for each mode if the defect layer thickness is increased and, after reaching a minimum, it increases a little then it reaches its second minimum and after this it sharply increases. The dependence of $A_{\max}$ on $d_1$ has analogous character, but here the central peak has an essential height, which means that one can obtain strong light absorption in the system (changing the defect layer thickness) for essentially low values of the parameter $n''$.

One can practically say the same thing if the absorbent is the CLC layer and the defect layer is non-absorbing, that is, for $n''_m \neq 0$, $n''_m = 0$ (see Fig. 5*b*). The picture is complicated for the gain case (Fig. 5*c,d*). If the dependences of $|n''|_{\max}$ and $|n''_m|_{\max}$ versus $d_1$ are practically the same as those of $(n'')_{\max}$ and $(n''_m)_{\max}$ versus $d_1$, then the dependence of $\ln|A|_{\max}$ versus $d_1$ is essentially different from that of $A_{\max}$ versus $d_1$. For $|n''| \neq 0$ and $|n''_m| = 0$ the value of $\ln|A|_{\max}$ has a strong minimum if the defect mode is practically in the centre of the PBG, that is, it is at the local maximum of $(n'')_{\max}$; and it has two maxima of significant height if the defect mode is displaced to the PBG borders. Nearby the PBG borders, $\ln|A|_{\max}$ sharply decreases. This means that here a laser generation is possible if the defect modes are not in the centre of the system but get closer to the PBG borders. Nevertheless, if these modes are essentially close to the PBG borders the low-threshold laser generation is again lost. For $|n''| = 0$ and $|n''_m| \neq 0$, the number of maxima of $\ln|A|_{\max}$ increases about two times.

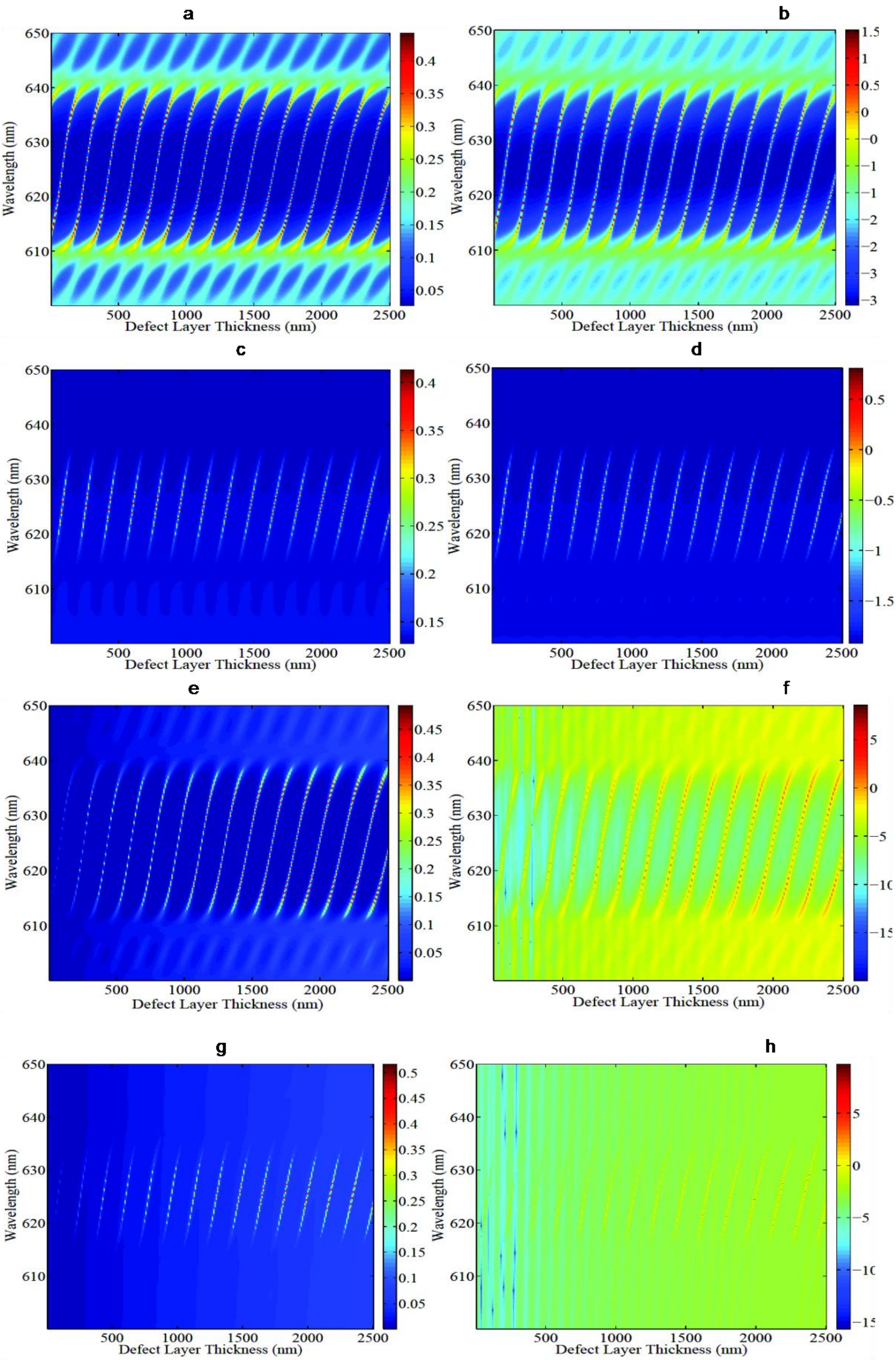

Fig.6. (Color online) The evolution (*a,c,e,g*) of the absorption spectra and (*b,d,f,h*) of the emission spectra when the defect layer thickness changes for (*a,b,e,f*) the diffracting and (*c,d,g,h*) non-diffracting circular polarizations. (*a,b,c,d*) correspond to case if $n_o^{''} = n_e^{''} \neq 0$ and $n^{''} = 0$, and (*e,f,g,h*) correspond to the case if $n_o^{''} = n_e^{''} = 0$ and $n^{''} \neq 0$.

In Fig. 6 the evolution (*a,c,e,g*) of the absorption spectra and (*b,d,f,h*) of the emission spectra when the defect layer thickness changes are presented for (*a,b,e,f*) the diffracting and (*c,d,g,h*) non-diffracting circular polarizations. (*a,b,c,d*) correspond to case if $n_o'' = n_e'' \neq 0$ and $n'' = 0$, and (*e,f,g,h*) correspond to the case if $n_o'' = n_e'' = 0$ and $n'' \neq 0$. As it was mentioned above, the defect mode wavelength increases from a minimum wavelength to a maximum of that of the band gap if the defect layer optical thickness increases. As we can see in this figure, for this change of the defect layer thickness, absorption and emission on the defect mode do not change smoothly, but with strong oscillations. And this means that there exists a definite value of the defect layer thickness at which the high absorption and emission takes place.

CLCs with defect inside as laser cavities have been the focus of intense research in recent decades. Lasing in these structures takes place due to defect modes, where spontaneous emission is inhibited in defect modes (where the photonic density of states (PDS) is large). The investigation of the PDS $\rho$ is important because of the following.

For laser emission it has been shown that, for instance, analyzing the case of the Fabry-Perot resonator, the threshold gain $g_{th}$ can be directly related to the maximum PDS $\rho_{\max}$, that is, $g_{th} \propto n/\rho_{\max} d$, where $n$ is the refractive index inside the resonator of the length $d$ [25]. Furthermore, according to the space-independent rate equations, the slope efficiency of lasers can be shown to be inversely proportional to the threshold energy and, therefore, directly proportional to $\rho_{\max}$ [26]. The PDS, i.e. the number of wave vectors $k$ per unit frequency – $\rho(\omega) = dk/d\omega$ – is the reverse of the group velocity and can be defined by the expression [26]:

$$\rho_i(\omega) \equiv \frac{dk_i}{d\omega} = \frac{1}{d}\,\frac{\dfrac{du_i}{d\omega}v_i - \dfrac{dv_i}{d\omega}u_i}{u_i^2 + v_i^2}, \quad i = 1,2 \tag{2}$$

where $d$ is the whole system thickness; ω is the incident light frequency; and $u_i$ and $v_i$ are the real and imaginary parts of the transmission amplitudes; $t_i(\omega) = u_i(\omega) + jv_i(\omega)$ are the transmission amplitudes for incident light with the two EPs; $j$ is the imaginary unit. For the isotropic case we have: $\rho_{iso} = n_s/c$, where $n_s = \sqrt{\varepsilon_s}$ is the refractive index of the media surrounding the system and $c$ is the speed of light in vacuum.

One must realize that the concept of the PDS introduced by Eq. (2) is not without controversy, especially in the case of non-periodic systems (more details about this see in [27]). Let us only note that although in general it is impossible to ascribe a direct physical meaning to the PDS in Eq. (2), anyhow, it can be used as a parameter providing some heuristic guidance in experiments for the PDS dispersion-related effects. Moreover, as it was showed in [28-31] Eq. (2) is applicable for finite PCs layers, even for the cases with absorption and amplification (naturally, for weak ones). In [32] it was showed that this equation is applicable (at least qualitatively) to the CLC layer with an isotropic layer (the aperiodic system) both in the presence of isotropic absorption and gain.

In [29-31], it was shown that in the absence of absorption, the PDS was proportional to the integral of the energy stored in the medium. Then, in [29, 30], it was shown that it takes place both in the presence of absorption and gain. Below we show that Eq.(2) is applicable (qualitatively) to the CLC layer with an isotropic defect layer (the aperiodic system) in the presence of anisotropic amplification in the CLC layer. To do this we calculate and compare the spectra of the PDS and the total field intensity $|E|^2$ aroused at the left border of the defect layer.

In Fig. 7 the evolution (*a, c*) of the relative PDS $\rho_1/\rho_{iso}$ spectra and (*b, d*) the total field intensity $|E|^2$ spectra when the gain parameter $x' = \ln(-2\,\mathrm{Im}\,\varepsilon_m)$ increases are presented for the diffracting eigen polarization. (*a, b,*) correspond to case if $n_e'' \neq 0$ and $n_o'' = 0$, and (*c, d*) correspond to the case if $n_e'' = 0$ and $n_o'' \neq 0$.

The presented results show that for sharp changes of $|E|^2$ and, consequently, also for the energy stored in the system, if the parameter $x'$ increases, the expression $\rho_1 / \rho_{iso}$ also undergoes sharp changes versus the same parameters. This means that Eq. (1) can also describe laser peculiarities of periodic systems with a defect in their structure. At the same time this figures give information about laser peculiarities of CLC layers with defects inside them in the presence of gain (about these peculiarities see [33]).

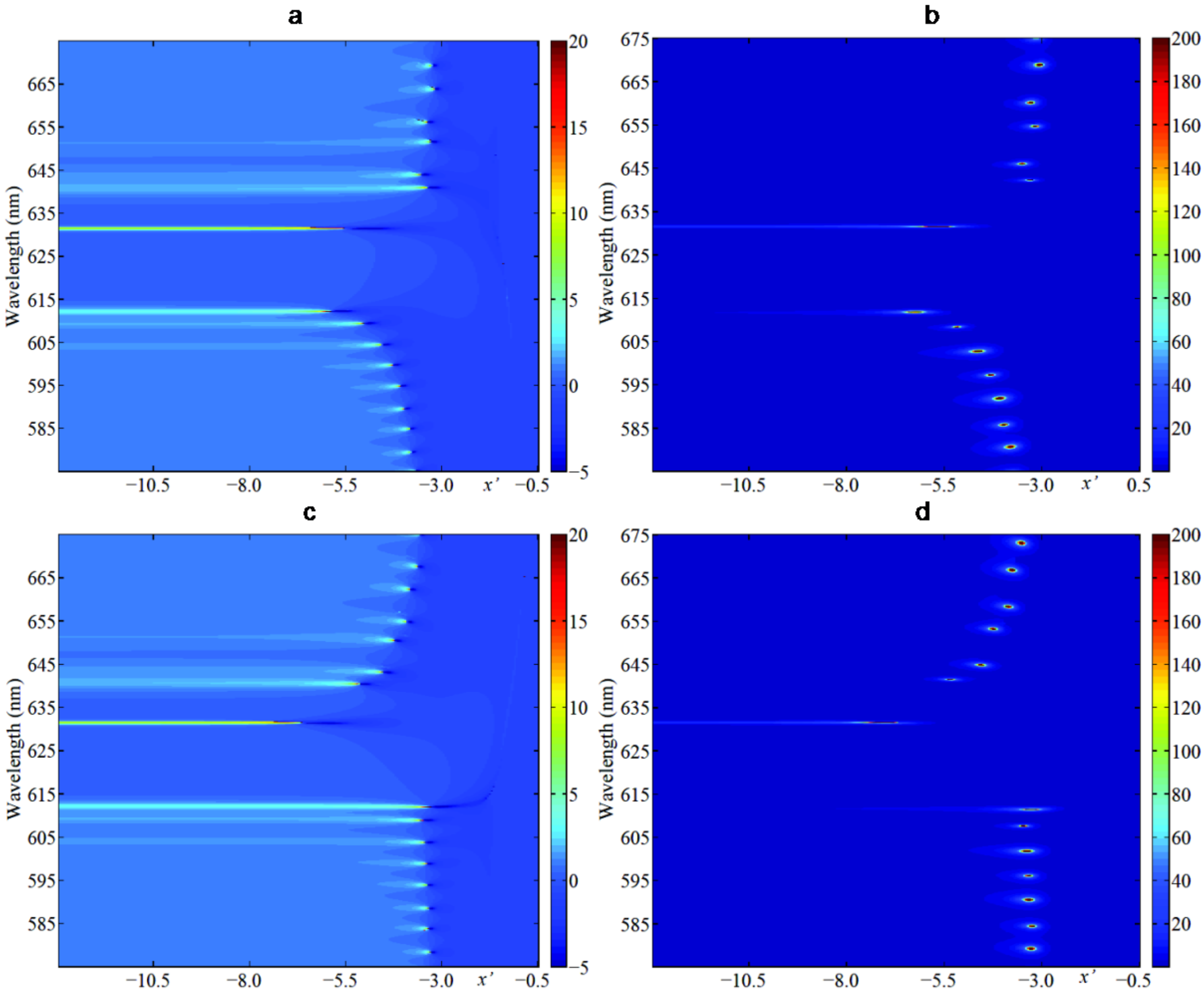

Fig.7. (Color online) The evolution (*a, c*) of the relative PDS $\rho_1 / \rho_{iso}$ spectra and (*b, d*) the total field intensity $|E|^2$ spectra if the gain parameter $x' = \ln(-2\,\mathrm{Im}\,\varepsilon_m)$, for the diffracting EP. (*a, b,*) correspond to the case if $n_e'' \neq 0$ and $n_o'' = 0$, and (*c, d*) correspond to the case if $n_e'' = 0$ and $n_o'' \neq 0$.

## 4. Conclusions

We investigated the influence of the defect layer location on absorption and emission of a CLC layer with an isotropic defect inside.

As our numerical calculations for non-diffracting circularly polarized incident light show, absorption and emission of the system are maximal if the defect is in the centre of the system; meanwhile, for the diffracting circularly polarized incident light absorption and emission are maximal if the defect is not in the centre of the system, but is a little displaced from it to the left border of the system (from where light is falling).

We found that for absorption the spectral line width is larger than for gain spectral line width for emission.

Then we investigated the absorption and emission dependences versus the parameters characterizing absorption and gain for different thicknesses of the defect layer.

We showed that the PDS defined in [26] applicable (qualitatively, at least) to the subject system, also is applicable for anisotropic absorption and gain in the CLC layer.

We also investigated the peculiarities of the total field intensity $|E|^2$ aroused in the defect layer for anisotropic absorption and gain.

Our results can be applied for designing of systems with strong absorption/emission, or having strong localization, or low-threshold laser generation etc.